\documentclass[12pt]{article}
\usepackage{amssymb}
\usepackage{latexsym}

\newif{\ifcomentarios}
\comentariosfalse
\newtheorem{theorem}{Theorem}
\newtheorem{lemma}[theorem]{Lemma}
\newtheorem{definition}[theorem]{Definition}
\newtheorem{proposition}[theorem]{Proposition}
\newtheorem{corollary}[theorem]{Corollary}

\newtheorem{remark}[theorem]{Remark}
\newtheorem{conjecture}[theorem]{Conjecture}

\newcommand{\calD}{{\cal D}}

\newcommand{\Z}{\mathbb Z}

\newcommand{\text}{\rm}
\newcommand{\dfrac}{\displaystyle\frac}
\newcommand{\dsum}{\displaystyle\sum}
\newcommand{\dprod}{\displaystyle\prod}
\newcommand{\dint}{\displaystyle\int}

\begin{document}

\author{{\bf Domingos H. U. Marchetti}\thanks{
e-mail: marchett{@}ime.usp.br}
\thinspace\ \& \thinspace {\bf Roberto da Silva}\thanks{
e-mail: dasilva{@}gibbs.if.usp.br} \\
Instituto de F\'\i sica \\
Universidade de S\~ao Paulo\\
P. O. Box 66 318\\
05315-970 S\~ao Paulo, SP, Brazil}
\title{Brownian Motion Limit of Random Walks in Symmetric Non--Homogeneous Media}
\date{}
\maketitle

\begin{abstract}
The phenomenon of macroscopic homogenization is illustrated with a simple
example of diffusion. We examine the conditions under which a
$d$--dimensional simple random walk in a symmetric random media
converges to a
Brownian motion. For $d=1$, both the macroscopic homogeneity condition and
the diffusion coefficient can be read from an explicit expression for the
Green's function. Except for this case, the two available formulas for the
effective diffusion matrix $\kappa $ do not explicit show how macroscopic
homogenization takes place. Using an electrostatic analogy due to Anshelevich,
Khanin and Sinai \cite{AKS}, we discuss upper and lower bounds on the
diffusion coefficient $\kappa $ for $d>1$.
\end{abstract}


\section{Introduction}

The long time behavior of random walks on a random environment is reviewed.
We focus mainly on the following question:

{\em What are the conditions under which a properly scaled random walk on a
non--homogeneous medium converges to a Brownian motion}.

This and related phenomena are usually named macroscopic homogenization (and
the environmental conditions are called {\it macroscopic homogeneity
conditions)} because such system looks homogeneous at macroscopic scales.
The discussion will be restricted to simple random walks with non--vanishing
transition probabilities (or rates) $\{w_{\langle xy\rangle }\}$ such that 
\begin{equation}
w_{\langle xy\rangle }=w_{\langle yx\rangle }  \label{w}
\end{equation}
holds for all nearest neighbor sites $\langle xy\rangle $ of a $d$
--dimensional lattice ${\Bbb Z}^{d}$. The so--called symmetric medium has
been considered by several authors (see e.g. \cite{ABSO,AKS,AV,MFGW,Ku,PV}
and references therein).

Except for a basic lemma, the general scheme of our presentation will be
dimensional independent. However, the one--dimensional problem plays a
central role in this work since, in this case, the macroscopic homogeneity
condition can be read from an explicit formula.

The $d=1$ case has been mostly investigated. The first mathematical results 
\cite{KKS,So,Si} were concerned with asymmetric random walks with transition
probabilities $w_{x,x+1}=1-w_{x,x-1}$, $x\in {\Bbb Z}$, being independent
and identically distributed (i.i.d.) random variables (note that $
w_{x,x+1}\not=w_{x+1,x}$). The trajectories $\{X(t),t>0\,|\,X(0)=x_{0}\}$ of
asymmetric random walks were shown to behave very anomalously. Symmetric
random walks began to be discussed in a series of papers (see e.g. \cite
{ABO,BSW,ABSO}) in connection with the problem of disordered chains of
harmonic oscillators (see \cite{LM} for an introduction and a selection of
reprints) and other problems in physics. Using Dyson's integral equation 
\cite{D}, these authors derived the following asymptotic behavior for the
trajectories: if 
\begin{equation}
{\Bbb E}w_{x,x+1}^{-1}<\infty \,,  \label{Ew}
\end{equation}
where ${\Bbb E}$ denotes the expectation value with respect to the i.i.d.
random variables $\{w_{x,x+1}\}$, then 
\[
{\Bbb E}X^{2}(t)\sim 2\left( {\Bbb E}w_{x,x+1}^{-1}\right) ^{-1}t\,,
\]
as $t\rightarrow \infty $, and the convergence of $X(t)/\sqrt{t}$ to the
Gaussian random variable with zero mean and variance $2\left( {\Bbb E}
w_{x,x+1}^{-1}\right) ^{-1}$ is implied. In addition, if condition
(\ref{Ew}) is violated, then ${\Bbb E}X^{2}(t)$ grows as $t^{\delta }$
with an exponent $\delta <1$ depending on the divergence of the
distribution at $w_{x,x+1}=0$.

A mathematical proof of convergence to Brownian motion for $d=1$ symmetric
random walks was given by Anshelevich and Vologodski \cite{AV}. For $d\geq 2$
there are at least two different proofs and both require macroscopic
homogeneity conditions more stringent than (\ref{Ew}). Anshelevich, Khanin
and Sinai \cite{AKS} proved the result by developing an expansion for the
expected value of the inverse of a non-homogeneous discrete Laplacian.
K\"{u}nnemann \cite{Ku} has proven this result by extending
Papanicolaou--Varadhan's approach \cite{PV}. Whether (\ref{Ew}) is a
necessary (and sufficient) macroscopic homogeneity condition for $d\geq 2$,
remains, to our knowledge, an open problem.

The present paper is inspired by the work of Anshelevich, Khanin and Sinai.
We use the logic of this proof in order to simplify the Anshelevich and
Vologodski's proofs for $d=1$. Our proof, in particular, eliminates the
technical hypothesis of $w_{x,x+1}$ to be strictly positive and illustrates
with textbook's mathematical methods the macroscopic homogeneity
condition (\ref{Ew}).

A Brownian motion is described by the diffusion equation. The random
environment induces an effective diffusive matrix $\kappa $ whose elements
are given by $\lim_{t\rightarrow \infty }{\Bbb E}\left(
X_{i}(t)X_{j}(t)\right) /t$, $i,j=1,\ldots ,d$. For the one--dimensional
problem, the reciprocal of this constant is given by the macroscopic
homogeneity condition: $\kappa ^{-1}={\Bbb E}w_{x,x+1}^{-1}$. For $d\geq 2$,
the two available formulas for the effective diffusion matrix (see \cite
{AKS,Ku}) do not explicitly show how macroscopic homogenization takes place
and this makes it difficult to obtain estimates. In this review (see also 
\cite{AKS}) the upper bound 
\[
\kappa \leq {\Bbb E}w_{\left\langle xy\right\rangle } 
\]
will be shown to hold if $\left| 1-w_{\left\langle xy\right\rangle }/{\Bbb E}
w_{\left\langle xy\right\rangle }\right| \leq \delta <1/2$. We also discuss
how a lower bound can be obtained using the electrostatic equivalence of the
diffusion problem as formulated in \cite{AKS}.

The outline of the present work is as follows. In Section \ref{results} we
formulate the problem and state our results. The proofs will be given in
Section \ref{RP} under the assumption that the eigenvalues and eigenvectors
of the semi--group generator of the process converge to the eigenvalues and
eigenfunctions of a Laplacian. The eigenvalue problem is a consequence of
our basic lemma (Lemma \ref{Green}) which will be proven in Section \ref{bl1}
for $d=1$ by Green\'{}s function method. The spectral perturbation
theory will be presented in
Section \ref{eigenvv}. Finally, the diffusion coefficient will be examined
in Section \ref{Difusion}.

\section{Statement of Results}

\label{results}

Let ${\Bbb B}$ denote the set of bonds of the regular lattice ${\Bbb Z}^{d}$
and let $w=\{w_{b}\}_{b\in {\Bbb B}}$ be an assignment of positive numbers.
Each component $w_{b}$ represents the transition rate of a random walk to go
from the site $x$ to $y$ along the bond $b=\langle xy\rangle $. The
assignment $w$ defines an symmetric environment on ${\Bbb Z}^{d}$ if the
transition rates satisfy $w_{\langle xy\rangle }=w_{\langle yx\rangle }\,$.

Given an environment $w$ and a finite set $\Lambda \subset {\Bbb Z}^{d}$, a
continuous time random walk on $\Lambda $, with absorbing boundary
condition, is a Markov process $\{X_{\Lambda ,w}(t),\,t\geq 0\}$ with
differential transition matrix $W_{\Lambda }=W_{\Lambda }(w)$ defined by 
\begin{eqnarray}
\left( W_{\Lambda }\,u\right) _{x} &=&\dsum_{{{y\in {\Bbb Z}^{d}: \atop
|x-y|=1}}}w_{\langle xy\rangle }\,u_{y}  \nonumber \\
& = &\left( \Delta _{\Lambda ,w}\,u\right) _{x}+u_{x}\;\dsum_{{{
y\in {\Bbb Z}^{d}: \atop |x-y|=1}}}w_{\langle xy\rangle }\,,  \label{W_N}
\end{eqnarray}
where 
\[
\left( \Delta _{\Lambda ,w}\,u\right) _{x}=\dsum_{{{y\in {\Bbb Z}^{d}:
}\atop {|x-y|=1}}}w_{\langle xy\rangle }\,\left( u_{y}-u_{x}\right) \,, 
\]
for all $x\in \Lambda $ and $u$ such that $u_{y}=0$ if $y\in {\Bbb Z}
^{d}\setminus \Lambda $.

Note that $-\Delta _{\Lambda ,w}$ is a positive matrix, 
\begin{equation}
(u,(-\Delta _{\Lambda ,w})u)=\frac{1}{2}\sum_{{{x,y\in {\Bbb Z}^{d}:\atop
|x-y|=1}}}w_{\langle xy\rangle }\,\left( u_{y}-u_{x}\right) ^{2}\geq 0\,,
\label{positive}
\end{equation}
and, if $w_{b}=\overline{w}$ for all $b\in {\Bbb B}$, $\Delta _{\Lambda ,
\overline{w}}=\overline{w}\,\Delta _{\Lambda }$ where $\Delta _{\Lambda }$
is the finite difference Laplacian with $0$--Dirichlet boundary condition on 
$\Lambda $. From here on, $\Lambda $ is taken to be the hypercube centered
at origin with size $|\Lambda |=(2L-1)^{d}$, $L\in {\Bbb N}$: $\Lambda
_{L}:=\left\{ x=(x_{1},\dots ,x_{d})\in {\Bbb Z}^{d}:\sup_{i}|x_{i}|<L\right
\} $, and all quantities depending on $\Lambda $ will be indexed by $L$
instead.

The probability distribution of $\{X_{L,w}(t),\,t\geq 0\}$ is governed by
the semi--group $\widehat{{T}}_{t}^{L}$ generated by $\Delta _{L,w}$. If $
X_{L,w}(t)$ denotes the position of a random walk at time $t$, then 
\begin{equation}
P(X_{L,w}(t)=x|X_{L,w}(0)=0)=\left( e^{t\,\Delta _{L,w}}u_{0}\right) _{x}\,,
\label{ivp}
\end{equation}
where $(u_{0})_{x}=\delta _{0,x}$.

The semi--group $\widehat{{T}}_{t}^{L}$ is the solution of the initial
value problem in ${\Bbb R}^{\Lambda }$, 
\begin{equation}
\frac{du}{dt}-\Delta _{L,w}\,u=0\,,  \label{iveq}
\end{equation}
with initial condition $u(0)=u_{0}$.

The solution of (\ref{iveq}) exists for all times $t>0$ and all sizes $
L<\infty $ but may depend on the realization of $w$ and on the initial
value. We present the sufficient conditions on the environment $w$ by which
the solution of (\ref{iveq}), under suitable scaling of time and space,
converges almost everywhere in $w$ to the fundamental solution of the heat
equation, 
\begin{equation}
\frac{\partial u}{\partial t}-\partial ^{2}u=0\,,  \label{difusion}
\end{equation}
with $u(t,\partial {\cal D})=0$. Here, (\ref{difusion}) is defined in the
domain $t>0$ and $\xi \in {\cal D}:=(-1,1)^{d}$ with boundary $\partial 
{\cal D}=\{\xi :\sup_{i}|\xi _{i}|=1\}$, and $\partial ^{2}=\partial
^{2}(\kappa )$ is given by 
\begin{equation}
\partial ^{2}=\partial \cdot \kappa \partial =\dsum_{i,j=1}^{d}\kappa _{ij}
\dfrac{\partial ^{2}}{\partial _{\xi _{i}}\partial _{\xi _{i}}}\,.
\label{glap}
\end{equation}

The heat kernel $T_{t}(\eta ,\zeta )=e^{t\,\kappa \partial ^{2}}(\eta ,\zeta
)$ when defined in ${\Bbb R}^{d}\times {\Bbb R}^{d}$ gives rise to a Wiener
process (or Brownian motion) $\{B(t),\,t\geq 0\}$ with covariance ${\Bbb E}
B(t)B(s)=4\kappa \,{\rm min}(s,t)$ (see e.g. Simon \cite{S}). In view of the
boundary condition, $T_{t}$ yields a Brownian motion $\{B_{0}(t),\,t\geq 0\}$
on the domain ${\cal D}$ with absorbing frontiers. The probability density
of $B_{0}(s+t)-B_{0}(s)$ to be equal to $\xi $ can be obtained by solving
equations (\ref{difusion}) by Fourier method 
\begin{equation}
T_{t}(\xi ,0)=\sum_{n\in {\Bbb N}_{+}^{d}}\,e^{{-}(\pi
^{2}/4)\,n^{2}t}\prod_{i=1}^{d}{\rm cs}\left( \pi n_{i}\xi _{i}/2\right)
\,\,,  \label{Poisson}
\end{equation}
where $n^{2}=n\cdot \kappa n={\sum_{i,j=1}^{d}}n_{i}\kappa _{ij}n_{i}$ and $
{\rm sc}(n_{i}x)$ stands for either $\cos \left( n_{i}x\right) $ or $\sin
\left( n_{i}x\right) $ depending on whether $n_{i}$ is an odd or even
number.

This discussion suggests the following definition.

\begin{definition}
\label{BMconv} The random walk in a random environment $w$ is said to
converge to a Brownian motion if there exist a constant $\kappa
=\kappa (w)$, the diffusion coefficient, such that 
\[
\lim_{L\rightarrow \infty }\left( e^{L^{2}\,t\,\Delta _{L,w}}u_{0}\right)
_{[L\xi ]}=\left( e^{t\,\partial ^{2} (\kappa )}\,\mu _{0}\right) (\xi )
\]
uniformly in $t>0$, $\xi \in {\cal D}$ and $\mu _{0}$ on the space of finite
measures ${\cal M}$ with support in ${\cal D}$. For $r\in {\Bbb R}^{d}$, $
[r]\in {\Bbb Z}^{d}$ and has as components the integer part of the
components of $r$; and $u_{0}\in {\Bbb R}^{\Lambda _{L}}$ is a vector given by 
\[
(u_{0})_{x}=L^{d}\int_{x_{1}/L}^{(x_{1}+1)/L}\cdots
\int_{x_{d}/L}^{(x_{d}+1)/L}d\mu _{0}(\zeta )\,.
\]
\end{definition}

It is important to note that, by the dominant convergence theorem, this
definition implies the convergence in distribution of the random walk
process $\left\{ (1/L)X_{L,w}(L^{2}t),\,t\geq 0\right\} $ to the Brownian
motion $\{B_{0}(t),\,t\geq 0\}$ as it is known in Probability Theory (weak
convergence of their distributions): ${X}_{L}(t)\equiv \dfrac{1}{L}%
X_{L,w}(L^{2}t)\longrightarrow B_{0}(t)$ in distribution if, for any
collection $0<s_{1}<\cdots <s_{n}$ of positive numbers and any collection $
f_{1},\dots ,f_{n}$ of bounded and continuous functions in ${\cal D}$, $n\in 
{\Bbb N}$, we have 
\begin{equation}
{\Bbb E}_{\mu _{0}}f_{1}({X}_{L}(s_{1}))\cdots f_{n}({X}_{L}(s_{n}))
\longrightarrow {\Bbb E}_{\mu _{0}}f_{1}({B}_{0}(s_{1}))\cdots
f_{n}(B_{0}(s_{n}))\,,  \label{weakconv}
\end{equation}
as $N\rightarrow \infty $, where ${\Bbb E}_{\mu _{0}}$ means the expectation
of the process starting with the measure $\mu _{0}$. Note also that ${X}
_{L}(t)$ has been scaled as in the central limit theorem: ${X}_{L}(t)$ is a
sum of about $L^{2}$ independent increments\footnote{%
A simple random walk with continuum time jumps according to a Poisson
process on time with rate $2d$ and there will be $2dL^{2}t$ jumps in average
after a time $L^{2}t$. With the random environment, the Poisson process has
a site dependent rate given by $\sum_{y:|x-y|=1}w_{\left\langle
xy\right\rangle }$.} divided up by $L$.

\begin{theorem}[Anshelevich and Vologodski \protect\cite{AV}]
\label{rwtobm1} If $d=1$ and the environment $w$ is a stationary process
such that the partial sums 
\begin{equation}
s_{x}(w)=\frac{1}{x}\sum_{z{=0}}^{x-1}\frac{1}{w_{z,z+1}}  \label{sn}
\end{equation}
converge as $x\rightarrow \infty $ to $\kappa ^{-1}$, $0<\kappa <\infty $,
almost everywhere in $w$, then a random walk in a random environment $w$
converges, for almost every $w$, to a Brownian motion with diffusion
coefficient $\kappa $.
\end{theorem}

\begin{theorem}[Anshelevich, Khanin and Sinai \protect\cite{AKS}]
\label{rwtobm} For any $d\geq 1$ and $\delta <1/2$, let $w=\{w_{b}\}_{b\in 
{\Bbb B}}$ be independent and identically distributed random variables such
that 
\begin{equation}
\left| 1-\frac{w_{b}}{{\bar{w}}}\right| \leq \delta \,,  \label{mhc}
\end{equation}
with ${\bar{w}=}{\Bbb E}w_{b}$. Then a random walk in a random environment $w
$ converges, for almost every $w$, to a Brownian motion with diffusion
coefficient matrix $\kappa =\kappa (d,\delta ,{\bar{w}})$.
\end{theorem}

\begin{remark}
\label{environ}
\begin{enumerate}
\item  Theorem \ref{rwtobm1} was proven in \cite{AV} under the
condition (\ref{sn}) with $w_{x,x+1}>0$. The positive condition has
been eliminated in our proof. Condition (\ref{sn}) is met if
$w=\{w_{b}\}$ are i.i d. random
variables with $0<{\Bbb E}w_{b}^{-1}<\infty $. Finiteness of first negative
moment seems to be, according to arguments presented in \cite{ABSO} (see
also \cite{FIN}), a sufficient and {\it necessary} homogeneity condition
since, otherwise, the walk would spend a extremely large time between jumps
leading the process to be sub--diffusive.

\item  Whether the homogeneity condition $0<{\Bbb E}w_{b}^{-1}<\infty $ is
also sufficient for $d>1$ is, to our knowledge, an open problem. It would
already be an important achievement to prove Theorem \ref{rwtobm} for any
distribution whose support is ${\Bbb R}$. It is unfortunate that both proofs
(see \cite{AKS,Ku}) require as a technical step $w_{b}$ to be bounded away
from $0$ and $\infty $.

\item  Theorem \ref{rwtobm} holds also if the transition probabilities $
\{w_{xy}\}$ do no vanish for $y-x$ belonging to a finite set ${\mathfrak U}^{+}$
that is able to generate ${\Bbb Z}^{d}$ by translations. Under the same sort
of condition (\ref{mhc}), \cite{AKS} have shown convergence to Brownian
motion satisfying the diffusion equation (\ref{difusion}).

\item  Theorem \ref{rwtobm} can also be extended to Brownian motion on $
{\Bbb R}^{d}$ if \ one combines the result with both absorbing and periodic
boundary conditions on $\partial {\cal D}$ (see details in \cite{AKS}).
\end{enumerate}
\end{remark}

\begin{theorem}
\label{difusion-k} Under the conditions of Theorem \ref{rwtobm} and a
conjecture formulated in (\ref{CPi}), there exist a finite constant
$C=C(d)$, such that the bounds 
\begin{equation}
1-\varrho \leq \frac{\kappa }{{\Bbb E}w_{b}}\leq 1  \label{uplo}
\end{equation}
hold with $1$ being the $d\times d$ identity matrix and $\varrho $ a
positive matrix such that 
\[
\varrho \leq \frac{4\delta C\left( 1+C\right) }{1-4\delta \left( 1+C\right) }
\,,
\]
in the sense of quadratic forms.
\end{theorem}



\section{Basic Lemma}

\label{RP}

The proof of Theorems \ref{rwtobm1} and \ref{rwtobm} presented in this
section are based on the uniform convergence of the eigenvalues and
eigenvectors of $\Delta _{L,w}$ to the eigenvalues and eigenfunctions of the
Laplacian operator $\partial ^{2}$ on ${\cal D}$. The uniform
convergence follows from a classical result in perturbation theory which
says the following.

If $A_{1},A_{2},\ldots, A_{n}, \dots $ is a sequence of bounded
operators in a Hilbert
space ${\cal H}$ which converge to $A$ in the operator norm, then all
isolated pieces of their spectrum and their respective projections converge
uniformly, as $n\rightarrow \infty $, to those of $A$.

Because $\Delta _{L,w}$ and $\partial ^{2}$ are unbounded operators
we consider their inverse instead. To formulate the results of this
section, we need some definitions.

Let $i_{L}$ be an {\it isometry} of the vector space ${\Bbb R}^{\Lambda }$
into the piecewise constant functions in the vector space $L_{0}^{2}({\cal
D})$, of square--integrable functions $f$ on ${\cal D}=(-1,1)^{d}$ with
$f(\partial {\cal D})=0$: 
\[
i_{L}:{\Bbb R}^{\Lambda }\longrightarrow L_{0}^{2}({\cal D}) \, ,
\]
given by
\begin{equation}
(i_{L}\, u)(\xi )=\left\{ 
\begin{array}{lll}
u_{[L\xi ]} & {\rm if} & \xi \in (-1,1)^{d} \\ 
0 & {\rm if} & \xi \in \partial {\cal D} \, ,
\end{array}
\right.  \label{iso}
\end{equation}
with $[x]$ as in Definition \ref{BMconv}.

The adjoint operator $i_{L}^{\dagger }:L_{0}^{2}({\cal D})\longrightarrow 
{\Bbb R}^{\Lambda }$,
\begin{equation}
(i_{L}^{\dagger }\,f)_{x}=L^{d}\int_{x_{1}/L}^{(x_{1}+1)/L}dx_{1}\cdots
\int_{x_{d}/L}^{(x_{d}+1)/L}dx_{d}\,f(x)\,,  \label{iso2}
\end{equation}
is defined by the equation 
\[
\left\langle f\,,i_{L}\,u\right\rangle =(i_{L}^{\dagger }\,f\,,u) \, ,
\]
with the inner product in ${\Bbb R}^{\Lambda }$ and $L_{0}^{2}({\cal D})$
given, respectively, by 
\begin{equation}
\left( u,v\right) =\frac{1}{L^{d}}\sum_{x\in \Lambda }u_{x}\,v_{x}
\label{inner1}
\end{equation}
and 
\begin{equation}
\left\langle f,g\right\rangle =\int_{[-1,1]^{d}}f(x)\,g(x)\,dx\,.
\label{inner2}
\end{equation}

We shall denote by 
\begin{equation}
\Xi _{L}^{-1}:= L^{-2}\left( i_{L}\,\Delta
_{L,w}^{-1}\,i_{L}^{\dagger }\right)  \label{XiL}
\end{equation}
the scaled inverse of $\Delta _{L,w}$. The operator kernel of $\Xi _{L}^{-1}$
is a step function with step--width $1/L$ which, as we shall see in the next
lemma, approximates the kernel $(\partial ^{2})^{-1}(r,s)$ in the
operator norm induced by $L_{0}^{2}$--norm: 
\begin{equation}
\Vert A\Vert :=\sup_{f:\Vert f\Vert _{2}=1}\Vert A\,f\Vert _{2} \, , \label{norm}
\end{equation}
where $\Vert f\Vert _{2}^{2}:=\left\langle f,f\right\rangle $.

\begin{lemma}[Basic Lemma]
\label{Green} Under the conditions of Theorems \ref{rwtobm} and
\ref{rwtobm}, there exist a {\bf finite} number $L_{0}=L_{0}(w)$ such
that, $\{\Xi _{L}^{-1};L>L_{0}\}$ is a sequence of bounded
self--adjoint operator in $L_{0}^{2}({\cal D})$ which converges 
\begin{equation}
\Vert \Xi _{L}^{-1}-(\partial ^{2})^{-1}\Vert \longrightarrow 0 \, ,
\label{limG}
\end{equation}
as $L\rightarrow \infty $, in the operator norm topology.
\end{lemma}

It thus follows from perturbation theory (see e.g. \cite{F}):

\begin{corollary}
\label{eigen} If $\lambda $ is an isolated eigenvalue of $(\partial
^{2})^{-1}$ and $E$ the orthogonal projection in its eigenspace ${\cal
E}$, one can find a subspace ${\cal E}^{L}\subset L_{0}^{2}({\cal
D})$, invariant under $\Xi _{L}^{-1}$, and a corresponding orthogonal 
projection $E^{L}$ such that 
\begin{equation}
\Vert E^{L}-E\Vert \longrightarrow 0\;\;\;\;\;\;\;\;\;{\rm and}
\;\;\;\;\;\;\;\;\;\left\| E^{L}\left( \Xi _{L}^{-1}-\lambda I\right)
E^{L}\right\| \longrightarrow 0 \, , \label{EE}
\end{equation}
as $L\rightarrow \infty $
\end{corollary}

Lemma \ref{Green} will be proven for $d=1$ random walks in
Section \ref{bl1}. This lemma reduces the Brownian motion limit
problem to the convergence of the inverse matrix $\Delta _{L,w}^{-1}$
to the inverse Laplacian $(\partial ^{2})^{-1}$ in the
$L_{0}^{2}$--operator norm topology. Corollary \ref{eigen} will be
proven in Section \ref{eigenvv}.

\medskip

\noindent {\it Proof of Theorems \ref{rwtobm1} and \ref{rwtobm} assuming
Corollary \ref{eigen}.} (As in Appendix 3 of \cite{AKS}) Let \break 
$\varphi :{\Bbb R}_{-}\longrightarrow {\Bbb R}$ be given by 
\begin{equation}
\varphi (\lambda )=e^{t/\lambda }\,  \label{phi}
\end{equation}
and note that $\varphi $ is uniformly continuous at $\lambda =0$ with $
\varphi (0)=0$. We have 
\[
T_{t}=\varphi ((\partial ^{2})^{-1})\hfill \hspace{0.75in}{\rm and}
\hspace{0.75in}T_{t}^{L}=e^{t\,\Xi _{L}}=\varphi \left( \Xi
_{L}^{-1}\right) \,. 
\]

In view of Definition \ref{BMconv} and the isometry $i_{L}$, Theorems \ref
{rwtobm1} and \ref{rwtobm} can be restated as\footnote{The isometry
$i_{L}$ has been introduced to bring all operators to the same
Hilbert space $L_{0}^{2}({\cal D})$. Note, however, that $\Xi _{L}^{-1}$ and 
$T_{t}^{L}$ remain finite rank operators.} 
\begin{equation}
\sup_{\mu _{0}\in {\cal M}}\sup_{\xi \in {\cal D}}\left| \left[
(T_{t}-T_{t}^{L})\mu _{0}\right] (\xi )\right| \longrightarrow 0  \, ,\label{thm}
\end{equation}
as $L \to \infty $. We shall prove an equivalent statement: 
\[
\lim_{L\rightarrow \infty }\Vert T_{t}-T_{t}^{L}\Vert =\lim_{L\rightarrow
\infty }\Vert \varphi ((\partial ^{2})^{-1})\hfill -\varphi \left(
\Xi _{L}^{-1}\right) \Vert =0\;.
\]

The inverse Laplacian $(\partial ^{2})^{-1}$ on ${\cal D}$ with $0$
--Dirichlet boundary condition is a compact operator with spectral
decomposition given by (recall equation (\ref{Poisson})) 
\begin{equation}
(\partial ^{2})^{-1}=\sum_{n\in {\Bbb N}_{+}^{d}}\lambda _{n}\,E_{n}
\, , \label{sd}
\end{equation}
where $\lambda _{n}=\left( (\pi /2)^{2}\dsum_{i,j=1}^{d}n_{i}\kappa
_{i,j} n_{j} \right) ^{-1}\equiv 4/(\pi ^{2}n^{2})$ and $e_{n}(\xi
)=\dprod_{i=1}^{d}{\rm sc} 
[(\pi /2)n_{i}\,\xi _{i}]$, $n\in {\Bbb N}_{+}^{d}$, are eigenvalues and
associate eigenfunctions of $(\partial ^{2})^{-1}$ and 
\begin{equation}
E_{n}f=\langle e_{n},f\rangle \,e_{n}\,.  \label{Proj}
\end{equation}

Because $0$ is an accumulation point, we introduce an integer cut--off $%
N<\infty $ and let 
\begin{equation}
(\partial _{N}^{2})^{-1}=\sum_{{{n\in {\Bbb N}_{+}^{d}:}\atop {|n|\leq N}}
}\lambda _{n}\,E_{n}\,.  \label{sdN}
\end{equation}
We have 
\begin{equation}
\Vert \varphi ((\partial _{N}^{2})^{-1})-\varphi ((\partial ^{2})^{-1})\Vert
^{2}=\left\| \sum_{{|n|>N}}\varphi (\lambda _{n})\,E_{n}\right\| ^{2}=\sum_{{
|n|>N}}\varphi ^{2}(\lambda _{n})  \, , \label{DD}
\end{equation}
which can be made as small as we wish by letting $N\rightarrow \infty $.
More generally, the uniform continuity of $\varphi $ at $0$ means the
following: given $\varepsilon >0$ and a non--positive bounded operators $A$,
we can find $\delta >0$ such that if $\Vert A\Vert <\delta $ we have $\Vert
\varphi (A)\Vert <{\varepsilon }/{3}$. We shall use this fact often in the
sequel.

From Corollary \ref{eigen}, there exist a projector 
\[
E^{L,N}=\sum_{{{n\in {\Bbb N}_{+}^{3}:}\atop {|n|\leq N}}}E_{n}^{L}\, , 
\]
where $E_{n}^{L}$ is, analogously to $E_{n}$, the projector on the invariant
subspace: $\Xi _{L}^{-1}{\cal E}_{n}^{L}={\cal E}_{n}^{L}$. Writing 
\begin{equation}
\Xi _{L,N}^{-1}=E^{L,N}\Xi _{L}^{-1}E^{L,N}\,  \label{EE1}
\end{equation}
and using the fact that $E^{L,N}$ is an orthogonal projector, we have 
\begin{equation}
\begin{array}{lll}
\varphi (\Xi _{L}^{-1})-\varphi (\Xi _{L,N}^{-1}) & = & \varphi \left( \Xi
_{L,N}^{-1}+(\Xi _{L}^{-1}-\Xi _{L,N}^{-1})\right) -\varphi (\Xi _{L,N}^{-1})
\\ 
& = & \varphi (\Xi _{L}^{-1}-\Xi _{L,N}^{-1})
\end{array}
\label{orth}
\end{equation}
(here $\varphi (A^{-1})$ is defined by its power series $I+tA+t^{2}A^{2}/2+
\cdots $).

We now show that $\Vert \Xi _{L}^{-1}-\Xi _{L,N}^{-1}\Vert $ can be made
smaller then $\delta =\delta (\varepsilon )$ if $L$ and $N$ are chosen
sufficient large. By Lemma \ref{Green}, there exist $L_{1}\geq L_{0}$, $
L_{1}=L_{1}(\delta )$, such that 
\begin{equation}
\Vert \Xi _{L}^{-1}-(\partial ^{2})^{-1}\Vert <\frac{\delta }{3}  \, ,
\label{D1}
\end{equation}
for all $L>L_{1}$. From (\ref{sd}) and (\ref{sdN}), there exist
$N_{1}=N_{1}(\delta )$ such that 
\begin{equation}
\Vert (\partial ^{2})^{-1}-(\partial _{N}^{2})^{-1}\Vert <\frac{\delta
}{3}\, ,
\label{DD1}
\end{equation}
if $N>N_{1}$. By Lemma \ref{Green}, there exist $L_{2}>L_{0}$, $
L_{2}=L_{2}(\delta )$, such that 
\begin{equation}
\Vert (\partial _{N}^{2})^{-1}-\Xi _{L,N}^{-1}\Vert <\frac{\delta }{3}
\label{DDD}
\end{equation}
holds for all $L>L_{2}$ with $N$ fixed.

If $L > \max (L_{1}, L_{2})$ and $
N> N_{1}$, equations (\ref{D1})--(\ref{DDD}) yield
\begin{equation}  \label{XX}
\|\Xi ^{-1}_{L}- \Xi ^{-1}_{L,N} \| \le \| \Xi ^{-1}_{L} - (\partial ^{2})
^{-1} \| + \| (\partial ^{2})^{-1} - (\partial ^{2}_{N})^{-1} \| +
\|(\partial ^{2}_{N}) ^{-1} - \Xi ^{-1}_{L,N} \| < \delta \, ,
\end{equation}
which implies, due the continuity of $\varphi$ and the orthogonal
relation (\ref{orth}), 
\begin{equation}  \label{XXP}
\| \varphi (\Xi _{L}) - \varphi (\Xi _{L,N}) \| < \frac{\varepsilon
}{3} \, .
\end{equation}

By uniform continuity of $\varphi$ and (\ref{DDD}), we also have 
\begin{equation}  \label{DDP}
\| \varphi ( (\partial ^{2}) ^{-1}) - \varphi ((\partial ^{2}_{N})^{-1}) \|
< \frac{\varepsilon }{3} \, .
\end{equation}

In addition, using the spectral decomposition of $(\partial ^{2}_{N})^{-1}$
and $\Xi _{L,N}^{-1}$, and taking into account 
\begin{equation}  \label{lElE}
\| \lambda _{n}E_{n} - \lambda _{n}^{L} E_{n}^{L} \| \le | \lambda _{n} -
\lambda _{n}^{L}| \| E_{n} \| + | \lambda _{n}^{L}| \| E_{n} - E_{n}^{L} \|
\end{equation}
and Lemma \ref{Green}, we can find $L_{3}> L_{0}$, $L_{3}=L_{3} (\varepsilon
,N)$, such that 
\begin{equation}  \label{DXP}
\| \varphi ((\partial ^{2}_{N}) ^{-1}) - \varphi (\Xi ^{-1}_{N,L}) \| < 
\frac{\varepsilon }{3} \, .
\end{equation}

Now, let $L>\max (L_{1},L_{2},L_{3})$. It then follows from
(\ref{XXP}), (\ref{DDP}) and (\ref{DXP}) 
\[
\begin{array}{lll}
\Vert \varphi ((\partial ^{2})^{-1})-\varphi (\Xi _{L}^{-1})\Vert & \leq & 
\Vert \varphi ((\partial ^{2})^{-1})-\varphi ((\partial _{N}^{2})^{-1})\Vert
\\ 
&  & \;\;\;\;\;\;\;\;\;\;+\,\Vert \varphi ((\partial _{N}^{2})^{-1})-\varphi
(\Xi _{N,L}^{-1})\Vert +\Vert \varphi (\Xi _{L}^{-1})-\varphi (\Xi
_{L,N}^{-1})\Vert <\varepsilon \, ,
\end{array}
\]
which implies strong convergence of the semi--group and completes the
proof of Theorem \ref{rwtobm}. \hfill $\Box $

\smallskip

\begin{remark}
\label{rmk} The introduction of the cut--off $N$ in (\ref{sdN}) is necessary
even for homogeneous environment. In this case, the eigenvalues $\lambda
_{n}^{L}$ and eigenvectors $e_{n}^{L}$ of $L^{-2}\,\Delta _{L}^{-1}$ can be
computed explicitly: 
\begin{equation}
\lambda _{n}^{L}=\left( 4L^{2}\sum_{i=1}^{d}\sin ^{2}\frac{\pi }{4L}
n_{i}\right) ^{-1}\,\hspace{0.5in}{\rm and\hspace{0.5in}}e_{n}^{L}(x)=
\prod_{i=1}^{d}{\rm sc}\left( \frac{\pi }{2L}n_{i}\,x_{i}\right)  \, ,\label{eL}
\end{equation}
with $n\in \Lambda ^{\ast }:=\{1,2,\dots (2L-1)\}^{d}$ (recall ${\rm sc}
\left( n_{i}x\right) $ stands for $\cos \left( n_{i}x\right) $ or $\sin \left(
n_{i}x\right) $, depending on whether $n_{i}$ is an odd or even
number). Note that $|\lambda _{n}-\lambda _{n}^{L}|$, with $\lambda _{n}$
given by (\ref{sd}), may not be small if $|n|=O(L)$. We always pick $N$ large
but fixed and let $L\rightarrow \infty $ in order (\ref{DXP}) to be true.
\end{remark}


\section{Proof of Lemma \ref{Green} for $d=1$}

\label{bl1} \setcounter{equation}{0} \setcounter{theorem}{0}

In this section we prove Lemma \ref{Green} for $d=1$. We consider the
second--order Sturm--Liouville difference equation 
\[
\Delta _{L,w}\,u= f  
\]
with $u (\partial {\calD })=0$, and use the method of Green to
calculate the matrix elements of $\Delta _{L,w}^{-1}$. 
This gives, in view of equation (\ref{XiL}), an explicit formula for the
operator kernel $\Xi _{L}^{-1}\left( r,s\right) $.

The procedure starts by looking for two linear independent solutions of the
homogeneous equation 
\begin{equation}
\left( \Delta _{L,w}\,u\right) _{x}=w_{\langle x-1,x\rangle
}(u_{x-1}-u_{x})-w_{\langle x,x+1\rangle
}(u_{x}-u_{x+1})=0 \, , \label{heq}
\end{equation}
with $x\in \{-L+1,\dots ,L-1\} $ and $u_{-L}=u_{L}=0$. Without loss of
generality, we set $w_{\langle L-1,L\rangle }=w_{\langle
-L,-L+1\rangle }=\kappa $.

\begin{proposition}
\label{prop} Let $\xi _{L}\in {\Bbb R}^{2L-1}$ be a vector valued function
of the environment $w$ given by 
\begin{equation}
(\xi _{L})_{x}= \eta _{L}\dsum_{y=-L+1}^{x}w_{\langle y-1,y\rangle
}^{-1}  \, , \label{xi}
\end{equation}
for all $x\in \{-L+1,\dots ,L-1\}$, where
\[
\eta _{L}^{-1} = \sum_{y=-L+1}^{L}w_{\langle y-1,y\rangle }^{-1} \, .
\]
Then $u_{1}=\xi _{L}$ and $u_{2}=1-\xi
_{L}$ are two linear independent solutions of (\ref{heq}).
\end{proposition}

\noindent {\it Proof.} $u_{1}=\xi _{L}$ is a solution of (\ref{heq}) by
simple verification and the same can be said of $u_{2}= 1-\xi _{L}$. For
this, note that 
\begin{equation}  \label{nabla}
w_{\langle x-1,x \rangle} (\nabla \xi_{L})_{x} = \eta _{L}
\end{equation}
holds uniformly in $x$, where $(\nabla u)_{x}= u_{x} - u_{x-1}$. It thus
remains to verify that they are linear independent.

Let $W=W(u_{1},u_{2};x)$ be the ``Wronskian'' of the two solutions $u_{1}$
and $u_{2}$ given by the following determinant 
\begin{equation}
W=\left| 
\begin{array}{cc}
(u_{1})_{x} & (u_{2})_{x} \\ 
w_{\langle x-1,x\rangle }(\nabla u_{1})_{x} & w_{\langle x-1,x\rangle
}(\nabla u_{2})_{x}
\end{array}
\right| \,.  \label{cW}
\end{equation}
It follows that two solutions $u_{1}$ and $u_{2}$ are linear independent if $
W(u_{1},u_{2};x)\not=0$ for all $x\in \{-L,\dots ,L\}$. Plugging
$u_{1}$ and $u_{2}$ into (\ref{cW}) we have, in view of (\ref{nabla}),
\begin{equation}
W=-\eta _{L}\,\left[ (\xi _{L})_{x}+1-(\xi _{L})_{x}\right] =-\eta
_{L} \, ,
\label{cW1}
\end{equation}
which concludes the proof of the proposition. \hfill $\Box $

\smallskip

The inverse matrix $\Delta _{L,w}^{-1}$ can be calculated by the so called
Green's function method (see e.g. \cite{J}): 
\begin{equation}
\left( \Delta _{L,w}^{-1}\right) _{x,y} = \left\{ 
\begin{array}{lll}
\dfrac{(\xi _{L})_{x}\left[ 1-(\xi _{L})_{y}\right] }{-\eta _{L}\,} & {\rm if
} & x\leq y \\ 
&  &  \\ 
\dfrac{\left[ 1-(\xi _{L})_{x}\right] \,(\xi _{L})_{y}}{-\eta _{L}\,} & {\rm 
if} & x>y \,.
\end{array}
\right.   \label{Gf}
\end{equation}

To see this is true, we note $\left( \Delta _{L,w}^{-1}\right) _{z,y}$ is
the $z$--component of a vector for each $y$ fixed. So, by definition 
\[
\left( \Delta _{L,w}\,\Delta _{L,w}^{-1}\right) _{x,y}=0 
\]
holds for all $x\not=y$. For $x=y$ we have 
\begin{eqnarray*}
\left( \Delta _{L,w}\,\Delta _{L,w}^{-1}\right) _{x,x} &=&w_{\langle
x,x+1\rangle }\left( \nabla \,\Delta _{L,w}^{-1}\right) _{x+1,x}-w_{\langle
x-1,x\rangle }\left( \nabla \,\Delta _{L,w}^{-1}\right) _{x,x} \\ 
&=&-w_{\langle x,x+1\rangle }(\nabla \xi _{L})_{x+1}\dfrac{(\xi
_{L})_{x}}{-\eta _{L}}-w_{\langle x-1,x\rangle }(\nabla \xi _{L})_{x}\dfrac{1-(\xi
_{L})_{x}}{-\eta _{L}} \\
&=&(\xi _{L})_{x}+(1-(\xi _{L})_{x})=1 \, ,
\end{eqnarray*}
by (\ref{nabla}), verifying the assertion.

\smallskip

We are now ready to write the operator kernel of $\Xi _{L}^{-1}$. In view of 
\begin{eqnarray}
(i^{\dagger }\,g,A\,i^{\dagger }\,f) &=&\dfrac{1}{L}\dsum_{x,y\in \Lambda
}(i^{\dagger }\,g)_{x}\,A_{x,y}\,(i^{\dagger }\,f)_{y}  \nonumber \\
&=&\dfrac{1}{L}\dsum_{x,y\in \Lambda
}L\int\nolimits_{x/L}^{(x+1)/L}dr\,g(r)\,A_{x,y}\,L\int\nolimits_{y/L}^{(y+1)/L}ds
\, g(s) \label{ii} \\
&=&\dint\nolimits_{-1}^{1}dr\dint\nolimits_{-1}^{1}ds\,g(r)\,\left(
L\,A_{[Lr],[Ls]}\right) \,f(s)\;\;=\;\;\langle g,i\,A\,i^{\dagger
}\,f\rangle \, ,
\nonumber
\end{eqnarray}
valid for any $(2L-1)\times (2L-1)$ matrix $A$ and $f\in
L_{0}^{2}({\cal D})$, and definitions (\ref{XiL}) and (\ref{iso}), we
have 
\begin{equation}
\Xi _{L}^{-1}(r,s)= L^{-2}\left( i\,\Delta _{L,w}^{-1}\,i^{\dagger
}\right) (r,s)= L^{-1}\left( \Delta _{L,w}^{-1}\right) _{[Lr],[Ls]}\, , 
\label{Gf-f}
\end{equation}
for any $-1\leq r,s\leq 1$.

If new variables  
\[
\zeta _{L}:=2\,\xi _{L}-1 
\]
are introduced into equation (\ref{Gf}), the operator kernel
(\ref{Gf-f}) can be written as  
\begin{equation}
\Xi _{L}^{-1}(r,s)=\frac{- 1}{4L\eta _{L}}\left\{ 1-\left| (\zeta
_{L})_{[Lr]}-(\zeta _{L})_{[Ls]}\right| -(\zeta _{L})_{[Lr]}\,(\zeta
_{L})_{[Ls]}\right\}  \, ,\label{Gff}
\end{equation}
in view of the fact that $(\zeta _{L})_{x}$ is a monotone increasing
function of $x$.
\medskip

By Schwarz inequality the operator norm (\ref{norm}) is bounded by the
$L^{2}$--norm of the operator kernel, the Hilbert--Schmidt norm $|\Vert K\Vert
|^{2}:=\dint_{-1}^{1}dr\dint_{-1}^{1}ds\,|K(r,s)|^{2}$. Since the functions
in the Hilbert space has compact support, we have 
\[
\left\| K\right\| \leq \left| \left\| K\right\| \right| \leq
4\sup_{-1<r,s<1}|K(r,s)| 
\]
and $\Vert \Xi _{L}^{-1}-(\partial ^{2})^{-1}\Vert \longrightarrow 0$
is implied by the pointwise convergence $\Xi _{L}^{-1}(r,s)\longrightarrow
(\partial ^{2})^{-1}(r,s)$ of the operator kernel. We shall see that
the latter convergence sense is consequence of the following result.

\begin{proposition}
\label{zeta-r} Given $\varepsilon >0$ and $w$ satisfying the hypothesis of
Theorem \ref{rwtobm1}. Then, there exist an integer number
$L_{0}=L_{0}(\varepsilon ,w)$ such that 
\begin{equation}
\left| (\zeta _{L})_{[Lr]}-r\right| <\varepsilon  \label{z-r}
\end{equation}
holds for all $L>L_{0}$ and $-1<r<1$.
\end{proposition}

\noindent {\it Proof.} Under the hypothesis of Theorem \ref{rwtobm1}
the strong law of large numbers holds and
\begin{equation}
2\,L\,\eta _{L}\longrightarrow \kappa  \, , \label{xi1}
\end{equation}
as $L\rightarrow \infty $, for almost every $w$ (see eq. (\ref{xi})).
Analogously, since $[Lr]/L\longrightarrow r$ as $L\rightarrow \infty $, we
have 
\begin{equation}
(\zeta _{L})_{[Lr]}=2\,L\,\eta _{L}\cdot \frac{\lbrack Lr]+L}{L}\cdot \dfrac{
1}{[Lr]+L}\dsum_{y=-L+1}^{[Lr]}w_{\langle y-1,y\rangle
}^{-1}-1\longrightarrow r  \, , \label{zz}
\end{equation}
for each $r\in (-1,1)$ and this gives (\ref{z-r}). \hfill $\Box $

\bigskip

The Green's function method can also be used to compute the integral kernel
of $(\partial ^{2})^{-1}$ as an operator in the Hilbert space
$L_{0}^{2}({\cal D})$. The two linear independent solutions of the
homogeneous equation 
\begin{equation}
\kappa \frac{d^{2}f}{dr^{2}}=0\;, \;\;\;\;\;\;\;\;-1<r<1 \, ,  \label{heq1}
\end{equation}
with boundary condition $f(-1)=f(1)=0$ are $f_{1}=1+r$ and $f_{2}=1-r$.
Replacing $u_{1(2)}$ and $w_{\langle x-1,x\rangle }(\nabla
u_{1(2)})$ by $f_{1(2)}$ and $\kappa (df_{1(2)}/dr)$ in
(\ref{cW}), gives $W=-2\,\kappa $. Substituting
these into (\ref{Gf}) following the simplification of (\ref{Gff}), yields 
\begin{equation}
(\partial ^{2})^{-1}(r,s)=\frac{-1}{2\kappa }\left\{ 1-|r-s|-r\,s\right\}
\label{Gf1}
\end{equation}
Note that $|(\partial ^{2})^{-1}(r,s)|\leq 1/ (2\kappa )$ and, in view
of (\ref{Gff}) and Proposition \ref{zeta-r}, 
\begin{equation}
\left| \Xi _{L}^{-1}(r,s)\right| \leq C  \label{boundness}
\end{equation}
holds uniformly in $r,s\in (-1,1)$ for all $L>L_{0}$.

Now, let $\rho _{L}(r):=(i\,\zeta _{L})(r)-r$ and ${\widehat{\Xi }}
_{L}^{-1}:=2L\,\eta _{L}\,\Xi _{L}^{-1}/\kappa $. Then, if $L>L_{0}$, in
view of (\ref{Gff}), (\ref{Gf1}) and Proposition \ref{zeta-r}, we have 
\begin{equation}
\left| \Xi _{L}^{-1}(r,s)-(\partial ^{2})^{-1}(r,s)\right| \leq \left| \frac{
2L\eta _{L}}{\kappa }-1\right| \left| {\Xi }_{L}^{-1}(r,s)\right| +\left| 
{\widehat{\Xi }}_{L}^{-1}(r,s)-(\partial ^{2})^{-1}(r,s)\right| \, ,
\end{equation}
where 
\begin{equation}
\begin{array}{lll}
\left| {\widehat{\Xi }}_{L}^{-1}(r,s)-(\partial ^{2})^{-1}(r,s)\right| & =
& \dfrac{1}{2\kappa }\left| \rho _{L}(r)\rho _{L}(s)+r\rho _{L}(s)+s\rho
_{L}(r)+|\rho _{L}(s)-\rho _{L}(r)|\right| \\ 
&  &  \\ 
& < & \dfrac {2 \varepsilon}{\kappa } +\dfrac{\varepsilon ^{2}}{2\kappa }
\end{array}
\end{equation}
uniformly in $r,s\in (-1,1)$. When combined with (\ref{xi1}) and (\ref
{boundness}) this proves Lemma \ref{Green} for $d=1$.  

\hfill $\Box $


\section{Perturbation of Spectra}
\label{eigenvv}

{\it Proof of Corollary \ref{eigen}.} This proof can be found in Appendix B
of \cite{AKS} and is essentially based on the perturbation theory of
Hermitian bounded operators developed by Friedrichs in \cite{F}. Since it
can be described shortly, we repeat the proof's derivation for completeness. Our
derivation, however, is more close to \cite{F} in the sense that we
perturb an interval of the spectrum. When the interval contains one single
eigenvalue this reduces to the derivation of \cite{AKS}. The generalization
is however essential in dealing with intervals containing accumulation point
of the spectrum. This situation has to be considered in order to show that
the spectrum projection in such intervals remains orthogonal when the
perturbation is turned on.

We now introduce some notation. Let $I_{0}\in {\Bbb R}$ be an isolated
closed interval of the spectrum $\sigma (\partial ^{-2})$ of $\partial
^{-2}$ defined with Dirichlet boundary condition on $\calD = (-1,1)$.
There exist $0<\delta <\infty $ and an interval $I\subset I_{0}$ such that $
I\cap \sigma (\partial ^{-2})=\sigma (\partial ^{-2})\cap I_{0}$ and 
\[
{\rm dist}\,\left( I_{0},{\Bbb R}\setminus I\right) >\delta \,. 
\]
Let ${\cal E}_{0}$ denote the eigenspace ${\rm span}\{e_{n}:\lambda _{n}\in
I_{0}\}\in L_{0}^{2}({\cal D})$ associated with $I_{0}$ and $E_{0}$ the
spectral projection onto ${\cal E}_{0}$. Let ${\cal E}_{1/L}$ denote the
subspace of the Hilbert space ${\cal H}:=L_{0}^{2}({\cal D})$ invariant
under the action of $\Xi _{L}$ and $E_{1/L}$ the projection (not necessarily
orthogonal) onto ${\cal E}_{1/L}$.

The projection $E_{1/L}$ is defined by the following set of equations: 
\begin{equation}  \label{EE0}
(I-E_{1/L}) \,\Xi ^{-1} _{L} \, E_{1/L} =0
\end{equation}
(i.e. ${\cal E} _{1/L}$ is an invariant subspace) and the two conditions 
\begin{equation}  \label{cEcE0}
E_{1/L} \, E_{0} = E_{1/L} \;\;\;\;\;\;\;\;\;\;\;\;\;\;\;\;\; {\rm and
}\;\;\;\;\;\;\;\;\;\;\;\;\;\;\;\;\;\; E_{0}\, E_{1/L} = E_{0} \, .
\end{equation}
Note that, under (\ref{cEcE0}) $E_{1/L}$ is a projector 
\[
E_{1/L}^{2} =
(E_{1/L} \,E_{0})E_{1/L} = E_{1/L} (E_{0} \,E_{1/L}) =E_{1/L} \, E_{0} =
E_{1/L} \, ,
\]
which is consistent with $E_{0}$ in the sense that $\lim_{L\to
\infty}E_{1/L} = E_{0}$.

We shall prove that, provided $V_{L}:= \Xi _{L}^{-1} - \partial ^{-2}$ is
bounded, $E_{1/L}$ depends analytically on $1/L$ and ${\cal E} _{1/L}$ tends
to ${\cal E}_{0}$ as $1/L \to 0$. The proof of this statement uses
equation (\ref{EE0}) to write an integral equation. For simplicity, we
shall drop the index $L$ of the quantities $\Xi ^{-1}_{L}$, $V_{L}$,
${\cal E} _{1/L}$ and $E_{1/L}$. 

Our stating point begin with equation 
\begin{equation}  \label{EDE}
\begin{array}{lll}
(1-E)\partial ^{-2} \, E & = & \partial ^{-2} \, E - E\partial ^{-2} \, E \\ 
& = & \partial ^{-2} \, E - E\partial ^{-2} \, ,
\end{array}
\end{equation}
which comes from the following facts. 

The operator $\partial ^{-2}$
commutes with the spectral projector $E_{0}$. Using this and 
equations (\ref{cEcE0}), we have 
\[
E\partial ^{-2} \, E = E \, E_{0} \partial ^{-2} \, E = E \, \partial ^{-2} \,
E_{0} \, E = E \, \partial ^{-2} E_{0} = E \, E_{0} \, \partial ^{-2} = E \,
\partial ^{-2} \, ,
\]
and this implies the second line of (\ref{EDE}). The commutation relation
$[\partial ^{-2}, E_{0}] =  0$ 
allows us to replace $E$ in the equation (\ref{EDE}) by $Q := E-E_{0}$ 
\begin{equation}  \label{EDE1}
(1-E)\partial ^{-2} \, E = \partial ^{-2} \, Q - Q \, \partial ^{-2}
\, .
\end{equation}

Combining (\ref{EE0}) with (\ref{EDE1}) and using $\Xi ^{-1}= \partial
^{-2} + V$, gives 
\begin{equation}  \label{EE00}
\begin{array}{lll}
\partial ^{-2} \, Q & = & Q \,\partial ^{-2} - (I- E) \, V \, E \\ 
& = & \partial ^{-2} \, Q - (I- E_{0}- Q) \, V \, (E_{0}+ Q) \, .
\end{array}
\end{equation}

Since the interval $I_{0}$ is isolate from the rest of the spectrum,
$\partial ^{-2}$ is an invertible bounded operator in the subspace
$(I-E_{0})\, {\cal H} $. We can solve the left hand side of
(\ref{EE00}) for $Q$ by defining 
\[
X \, f := \left\{ 
\begin{array}{lll}
\partial ^{2} \, f &  & {\rm if } \; f\in (I-E_{0})\, {\cal H} \\ 
0 &  & {\rm otherwise} \, .
\end{array}
\right. 
\]
Note that $X\, \partial ^{-2}=\partial ^{-2} \, X = I-E_{0} $ and
$\|X\| < \delta ^{-1}$.

Equation (\ref{EE00}) can thus be written as 
\begin{equation}  \label{g}
Q= g (Q) \, ,
\end{equation}
where 
\[
g (Q) = X \, \partial ^{-2} \, Q - X \, (I+ E_{0}- Q) \, V \,
(Q-E_{0})  \, .
\]

\begin{proposition}
\label{fix} The sequence $Q_{n},\,n=0,1,\dots $, of projectors defined by 
\begin{equation}
Q_{n}=g(Q_{n-1})  \, , \label{Qn}
\end{equation}
with initial condition $Q_{0}=0$ satisfies the conditions (\ref{cEcE0}) and
converges, $Q=\lim_{n\rightarrow \infty }Q_{n}$, to the unique solution of
equation (\ref{g}).
\end{proposition}

\noindent {\it Proof.} We have $\Vert Q_{n}\Vert \leq q<1$ for all $n\in 
{\Bbb N}$ provided $q$ is chosen small enough and $L$ is taken so large that
if $\Vert Q \Vert \leq q$ then 
\begin{equation}
\Vert g(Q)\Vert \leq \Vert X\Vert \,\Vert E_{0}\,\partial ^{-2}\Vert \,q+\delta
^{-1}\,(1+q)^{2}\Vert V_{L}\Vert \leq q \, . \label{Qq}
\end{equation}
Note that the smallness of $g$ depends on the smallness of $V$. Since 
\[
\Vert X\Vert \Vert E_{0}\,\partial ^{-2}\Vert \leq  \frac{|I_{0}|}{\delta
+|I_{0}|} :=\alpha <1 \, ,
\]
equation (\ref{Qq}) holds provided 
\begin{equation}\label{VL}
\Vert V \Vert \leq \frac{(1-\alpha )\delta}{(1+q)^{2}} \,q \, .
\end{equation}

Now, for fixed value of $q$, it can be shown (see ref. \cite{F} for details) 
\[
\left| g(Q)-g(Q^{\prime })\right| \leq \theta \left| Q-Q^{\prime }\right| 
\]
also holds with $\theta <1$ and this implies Proposition \ref{fix} by the
Banach fixed point theorem. \hfill $\Box $

\bigskip

We have proven the existence of a unique projector $E_{1/L}$ such that $
\|Q_{1/L}\|=\|E_{1/L}- E_{0}\|\le q $. Since $q$ can be made arbitrarily
small by taking $L$ sufficiently large so that (\ref{VL}) holds, we
have $\lim_{L\to \infty}\|E_{1/L}- E_{0}\| = 0$ and ${\cal E}
_{1/L}\longrightarrow {\cal E} _{0}$ as $L\to \infty$.

To complete the proof of Theorem \ref{Green}, we need to find an orthogonal
projector $E^{L}$ onto ${\cal E} ^{L} \equiv {\cal E} _{1/L}$ in order to
get (\ref{EE}). This is achieved by setting
\begin{equation}  \label{Ehat}
E^{L}= E_{1/L} (E_{1/L}^{\dagger}E_{1/L})^{-1}E_{1/L}^{\dagger} 
\end{equation}
and noting that the inverse operator $(E^{\dagger}E)^{-1}$ exist
because $\|Q\| \le q $ implies 
\[
\|E_{0}\, f\| \le \|E\, f\| + q \|f\| = \|E\, f\| + q \|E_{0} \, f\|
\, ,
\]
for any $f\in {\cal H} $ such that $E_{0}f = f$. As a consequence $\|E\, f\|
\ge (1-q) \|f\|$ and 
\[
\langle f,E^{\dagger}E \, f\rangle \ge (1-q)^{2} \|f\| \, .
\]

One can show, in addition, that $[\Xi ^{-1},E^{L}]=0$ for all $L$. Therefore,
for any $\lambda \in I_{0}\cap \sigma (\partial ^{-2})$, 
\begin{equation}
\begin{array}{lll}
\Vert E^{L}\left( \Xi ^{-1}-\lambda I\right) E^{L}\Vert & = & \Vert \left(
\Xi ^{-1}-\lambda I\right) E^{L}\Vert \\ 
& \leq & \Vert \left( \partial ^{-2}-\lambda I\right) \left(
E_{0}+(E^{L}-E_{0})\right) \Vert +\Vert V\,E^{L}\Vert \\ 
& \leq & \Vert \left( \partial ^{-2}-\lambda I\right) \Vert +\Vert \partial
^{-2}\Vert \,\Vert E^{L}-E_{0}\Vert +\Vert V_{L}\Vert \,.
\end{array}
\label{EXiE}
\end{equation}
Since the right hand side goes to zero as $L\rightarrow \infty $ this
concludes the proof of Corollary \ref{eigen}. \hfill $\Box $


\section{Diffusion Coefficient}

\label{Difusion} \setcounter{equation}{0} \setcounter{theorem}{0}

This section is devoted to the proof of Theorem \ref{difusion-k}. The
diffusion coefficient will be estimated throughout an expansion for the
expectation of the inverse matrix, ${\Bbb E}(\Delta _{L,w})^{-1}$, with $w$
satisfying the macroscopic homogeneity condition (\ref{mhc}). This is
justified in ref. \cite{AKS} in view of the fact that $(\Delta
_{L,w})^{-1}$, when properly scaled, converge to its expectation for
almost all environment $w$. Thus, the formula $\left( {\Bbb
E}(\Delta _{L,w})^{-1}\right) ^{-1}\sim \Delta _{L,\kappa  }$ is expect
to hold in the limit as $L\rightarrow \infty $. We will see that very
important cancellations take place by inverting the series expansion of
${\Bbb E} (\Delta _{L,w})^{-1}$.

A simple algebraic manipulation shows 
\begin{equation}
\left( -\Delta _{L,\overline{w}}\right) ^{1/2}\,\frac{1}{-\Delta _{L,w}}
\,\left( -\Delta _{L,\overline{w}}\right) ^{1/2}=\frac{1}{I-D_{L,w}}
\, , 
\label{1-D}
\end{equation}
where 
\begin{equation}
D_{L,w}:=\left( -\Delta _{L,\overline{w}}\right) ^{-1/2}\left( \Delta
_{L,w}-\Delta _{L,\overline{w}}\right) \left( -\Delta _{L,\overline{w}
}\right) ^{-1/2}  \label{D}
\end{equation}
is a well defined matrix since, in view of (\ref{positive}) and (\ref{mhc}), 
$-\Delta _{L,w}$ is positive and the square root of $-\Delta _{L,\overline{w}
}$ can be taken.

Choosing $\overline{w}={\Bbb E}\,w_{b}$ and use (\ref{W_N}) to write $\Delta
_{L,\overline{w}}=\overline{w}\Delta _{L}$ where $\Delta _{L}$ is the finite
difference Laplacian with $0$--Dirichlet boundary condition on $\Lambda $,
equation (\ref{D}) can be written as

\begin{equation}
D_{L,w}=\left( -\Delta _{L}\right) ^{-1/2}\Delta _{L,\alpha }\left( -\Delta
_{L}\right) ^{-1/2}  \, , \label{D-D}
\end{equation}
where $\alpha =\{\alpha _{b}\}$ given by $\alpha _{b}=w_{b}/\overline{w}-1$,
are i.i.d. random variables with mean 
${\Bbb E}\,\alpha _{b}=0$, such that 
\begin{equation}
|\alpha _{b}|\leq \delta <\frac{1}{2}   \label{alpha}
\end{equation}
holds in view of (\ref{mhc}). 


Equation (\ref{1-D}) suggests us the use of Neumann series to develop a
formal expansion of $(\Delta _{L,w})^{-1}$ in power of $D_{L,w}$ due
to the small parameter $\delta $. The remaining
of this section is devoted to the pointwise convergence of the matrix
element of $\left[ {\Bbb E}\,\left( I -D_{L,w}\right) ^{-1}\right] ^{-1}$.

Using (\ref{D-D}), we have 
\begin{equation}
\frac{1}{I-D_{L,w}}=I+\sum_{n\geq 1}\left( -\Delta _{L}\right) ^{-1/2}\Delta
_{L,\alpha }\left[ \left( -\Delta _{L}\right) ^{-1}\,\Delta _{L,\alpha }
\right] ^{n-1}\,\left( -\Delta _{L}\right) ^{-1/2}\,.  \label{1-D-1}
\end{equation}

To write (\ref{1-D-1}) in a more convenient form, let $\nabla _{L}:{\Bbb R}
^{\Lambda }\longrightarrow {\Bbb R}^{{\Bbb B}_{L}}$ be the finite difference
operator: 
\[
(\nabla _{L}\,u)_{\langle xy\rangle }=-(\nabla _{L}\,u)_{\langle yx\rangle
}=\sigma _{\langle xy\rangle }\left( u_{y}-u_{x}\right) \;. 
\]
where the sign $\sigma _{\langle xy\rangle }=\sum_{i}(y_{i}-x_{i})=\pm 1$,
according to whether $\langle xy\rangle $ is positively ($=1$) or negatively
($=-1$) oriented. $\nabla _{L}$ maps a $0$--form $u$ into a $1$--forms $
\nabla _{L}u$. Let $\nabla _{L}^{\ast }:{\Bbb R}^{{\Bbb B}
_{L}}\longrightarrow {\Bbb R}^{\Lambda }$ be its adjoint $\left( \omega
,\nabla _{L}\,u\right) =\left( \nabla _{L}^{\ast }\,\omega ,u\right) $, i.e.
the finite divergent operator which maps a $1$--form $\omega $ into a
$0$--form $\nabla _{L}^{\ast }\omega $ given by 
\[
(\nabla _{L}^{\ast }\,\omega )_{x}=\sum_{y:|x-y|=1}\omega _{\langle
yx\rangle }\, ,
\]
and let $M_{\alpha }:{\Bbb R}^{{\Bbb B}_{L}}\longrightarrow {\Bbb R}^{{\Bbb
B}_{L}}$ be the multiplication operator by $\alpha $: $(M_{\alpha }\,\omega
)_{b}:=\alpha _{b}\,\omega _{b}$.

With these notations, we have 
\begin{equation}
\Delta _{L,\alpha }=\nabla _{L}^{\ast }\,M_{\alpha }\,\nabla _{L} \, ,
\label{NMN}
\end{equation}
and its bilinear form reads 
\[
\left( u,\Delta _{L,\alpha }v\right) =\left( \nabla u,M_{\alpha }\,\nabla
v\right) =\frac{1}{L^{d}}\dsum_{b}\alpha _{b}\,(\nabla u)_{b}\,(\nabla
v)_{b} \, ,
\]
recovering expression (\ref{positive}) for the quadratic form.

Define 
\begin{equation}
\Phi :=\nabla _{L}\left( -\Delta _{L}\right) ^{-1}\nabla _{L}^{\ast }
\, ,
\label{Phi}
\end{equation}
and note that, since $\left( -\Delta _{L}\right) _{x,y}^{-1}$ is the
{\it Coulomb potential} between two unit charges located at $x$ and
$y$, $\Phi _{b,b^{\prime }}$ is the {\it dipole interaction potential}
between two unit dipoles located at $b$ and $b^{\prime }$. Note that
$\Phi $ maps $1$--form into $1$--form.

In view of (\ref{NMN}) and (\ref{Phi}), equation (\ref{1-D-1}) can be
rewritten as 
\begin{equation}
\frac{1}{I-D_{L,w}}=I+\sum_{n\geq 1}\sum_{\Gamma \in {\Bbb B}_{L}^{n}}\alpha
_{\Gamma }\,W_{\Gamma }  \, , \label{1D1}
\end{equation}
where, for $\Gamma =(b_{1},b_{2},\dots ,b_{n})$, 
\begin{equation}
\alpha _{\Gamma }=\prod_{k=1}^{n}\alpha _{b_{k}}  \, , \label{aG}
\end{equation}
and 
\begin{equation}
\left( v,W_{\Gamma }\,u\right) =\frac{1}{|\Lambda |}\,\vartheta
_{b_{1}}\,\Phi _{b_{1},b_{2}}\,\Phi _{b_{2},b_{3}}\,\cdots \,\Phi
_{b_{n-1},b_{n}}\,\nu _{b_{n}} \, , \label{WG}
\end{equation}
with $\vartheta $ and $\nu $ being $1$--forms given by $\nabla \left(
-\Delta _{L}\right) ^{-1/2}v$ and $\nabla \left( -\Delta _{L}\right)
^{-1/2}u $, respectively.

Concern the convergence, as $\Lambda \nearrow {\Bbb Z}^{d}$, of a generic
term of the expansion (\ref{1D1}), the following remark is now in order.

\begin{remark}
\label{log} The asymptotic behavior of the dipole potential $\Phi
_{b,b^{\prime }}$ for $L>>{\rm dist}\,(b,b^{\prime })>>1$ can be estimated
by its spectrum decomposition \footnote{Since we have not rescaled the
space $\Lambda \subset \Z^{d}$, it is
convenient to introduce a base $\{{\hat{e}}_{n}^{L}\}_{n\in \Lambda ^{\ast }}
$, $\Lambda ^{\ast }=\{1,\dots ,2L-1\}^{d}$, normalized with respect to the
scalar product $(\!(u,v)\!):=\sum_{x\in \Lambda
}u_{x}\,v_{x}= L^{d} (u,v)$: ${\hat{e}}_{n}^{L}={e}_{n}^{L}/\sqrt{
L^{d}}$. The spectrum resolution of the identity is written in terms of
this base.},
\begin{equation}
\Phi _{b,b^{\prime }}=\frac{1}{L^{d}}\sum_{n\in \Lambda ^{\ast }}\tilde{
\lambda}_{n}\,(\nabla e_{n}^{L})_{b}\,(\nabla e_{n}^{L})_{b^{\prime }}
\, ,
\label{Phibb}
\end{equation}
where 
\[
\tilde{\lambda}_{n}^{-1}=4\sum_{k=1}^{d}\sin ^{2}\frac{\pi }{4L}\,n_{k}
\]
and $e_{n}^{L}$ as in (\ref{eL}). If we take $b=\langle xx^{(i)}\rangle $
and $b^{\prime }=\langle yy^{(j)}\rangle $, where $z^{(k)}$ is a nearest site
of $z$ whose components are given by $z_{\ell }^{(k)}=z_{\ell }+\delta
_{k,\ell }$, and make a change of variables, $\varphi _{i}=(\pi /2L)n_{i}$,
$i=1,\dots ,d$, we have  
\begin{equation}
\lim_{L\rightarrow \infty }\Phi _{b,b^{\prime }}=\frac{1}{4\pi ^{d}}
\int_{[0,\pi ]^{d}}d^{d}\varphi \left( \dsum_{k=1}^{d}\sin ^{2}(\varphi
_{k}/2)\right) ^{-1}\,\nabla ^{i}\nabla ^{j}\prod_{k=1}^{d}\cos \left(
\varphi _{k}(x_{k}-y_{k})\right)   \, , \label{LPhi}
\end{equation}
where $\nabla ^{k}f(z)=f(z^{(k)})-f(z)$ is the difference operator in
the $k$--th direction. The $|x-y|>>1$ behavior of $\Phi _{b,b^{\prime
}}$ is given by restricting the integral (\ref{LPhi}) around a
$\varepsilon $--neighborhood of $0$ with $\varepsilon \,|x-y|=O(1)$: 
\begin{eqnarray}
\Phi _{b,b^{\prime }} &\sim &\dfrac{-1}{(2\pi )^{d}}\dint_{|\varphi |\leq
\varepsilon }d^{d}\varphi \,\dfrac{\varphi _{i}\,\varphi _{j}}{\varphi ^{2}}
\,\tan \left( \varphi _{i}(x_{i}-y_{i})\right) \,\tan \left( \varphi
_{j}(x_{j}-y_{j})\right) \dprod_{k=1}^{d}\cos \left( \varphi
_{k}(x_{k}-y_{k})\right)   \nonumber \\
&\sim &\dfrac{-1}{(2\pi )^{d}|x-y|^{d}}\dint_{|t|\leq O(1)}d^{d}t\,\dfrac{
t_{i}\,t_{j}}{t^{2}}   \label{ass} \\
&\sim &\dfrac{1}{[{\rm dist}\,(b,b^{\prime })]^{d}}  \, . \nonumber
\end{eqnarray}
\end{remark}

As a consequence, $\Phi _{b,b^{\prime }}$ is not summable in absolute value,
\[
\sum_{b^{\prime }\in {\Bbb B}_{L}}\left| \Phi _{b,b^{\prime }}\right| \sim
\log L \, ,
\]
and the uniform convergence with respect to $\Lambda $ of the $\Gamma
$--summation in (\ref{1D1}) requires cancellations due to the dipole
orientations (see ref. \cite{PPNM}).

We shall exhibit in the following another kind of cancellation due to the
inversion of the expected value of (\ref{1D1}).

Inverting the expectation of (\ref{1D1}) gives 
\begin{equation}
\left[ {\Bbb E}\,\left( I-D_{L,w}\right) ^{-1}\right] ^{-1}=I -\Theta
_{L} \, ,
\label{D1D}
\end{equation}
where 
\begin{equation}
\Theta _{L}=\sum_{k\geq 1}(-1)^{k+1}\sum_{n_{1},\dots ,n_{k}}\sum_{\Gamma
=(\Gamma _{1},\dots ,\Gamma _{k})}{\Bbb E}\,\alpha _{\Gamma _{1}}\,\cdots \,
{\Bbb E}\,\alpha _{\Gamma _{k}}\,W_{\Gamma } \, , \label{Theta}
\end{equation}
and $\Gamma \in {\Bbb B}_{L}^{n_{1}}\times \cdots \times {\Bbb B}
_{L}^{n_{k}} $, with $n_{i}\geq 1$. Note that $W_{\Gamma }=W_{\Gamma
_{1}}\cdots W_{\Gamma _{k}}$. 

To see how the $\log $--divergent terms in (\ref{D1D}) cancel out, it is
convenient to use graph--theoretical language. A graph $G$ consists of two
sets $(V,E)$: $V=\{v_{1},\dots ,v_{s}\}$ is the vertex set and
$E=\{e_{1},\dots ,e_{s^{\prime }}\}$ the connecting set of edges. To
each edge $e$ its assigned an ordered pair of vertices $\left( vv^{\prime
}\right) $ (its extremities) which are called adjacent if $v\not=v^{\prime }$;
otherwise $e$ is said to be a ``loop''. To the problem at our hand, we shall
identify the bonds $\{b_{1},\dots ,b_{n}\}$ as a the vertex set of a
graph $G$ whose connectivity is determined by the presence of
interactions $\Phi _{b,b^{\prime }}$.

Two graphs $G$ and $G^{\prime }$ are {\it isomorphic} (denoted $G\sim
G^{\prime }$) if there is a one--to--one correspondence between their
elements
which preserves the incidence relation. A {\it path} $\Gamma $ on $G$ is an
ordered sequence $\{v_{i_{0}},e_{i_{1}},\dots ,e_{i_{n}},v_{i_{n}}\}$ of
alternately vertices and edges of $G$ such that $e_{i_{k}}=\left(
v_{i_{k-1}}v_{i_{k}}\right) $ holds for each $k$; the edges
$\{e_{i_{1}},\dots ,e_{i_{n}}\}$ are the steps of the path and the
vertices $\{v_{i_{0}},\dots ,v_{i_{n}}\}$ are the points visited by
the path. $\Gamma $ 
may be identified with one of these ordered sets since it can be uniquely
determined by each of them. Two vertices $v,v^{\prime }\in V$ may be
connected by more than one path. A graph $G$ is said to be {\it connected}
if any two vertices $v,v^{\prime }$ can be joined by at least one path
$\Gamma $ on $G$. The {\it components} of a non--connected graph are
its maximum connected subgraphs. Given two vertices $v,v^{\prime }$,
the {\it disconnecting set of edges} is a set whose removal from the graph
$G$ destroys all paths between $v,v^{\prime }$. A {\it cut--set} is a minimal set
of edges the removal of which from a connected graph $G$ causes it to fall
into two components $G_{1},G_{2}$.

Turning back to equation (\ref{D1D}), one may interpret $\Gamma
=\{b_{1},\dots ,b_{n}\}$ as a set of vertices visited by a path. In
view of the fact that $\alpha _{b}$ has zero mean, we have 
\begin{equation}
{\Bbb E}\,\alpha _{\Gamma }\equiv \bar{\alpha}_{\Gamma }=0  \label{Ealpha}
\end{equation}
if there exist at least one bond $b_{i}$ which are not repeated in the
list $\Gamma =\{b_{1},\dots ,b_{n}\}$. 

The condition (\ref{Ealpha}) says that the path $\Gamma $ must visit each
vertex at least twice otherwise its contribution to (\ref{D1D}) vanishes.
The set of distinct bonds $V=\{b_{i_{1}},\dots b_{i_{s}}\}$ and edges
$E=\{\left( b_{1}b_{2}\right) ,\dots ,\left( b_{n-1}b_{n}\right) \}$
form a connected graph $G$ with even valency ${\cal V}(b)\geq 4$ for
each vertex $b\in G$. Graphs with this property will be called {\it
admissible graphs}.
Note that each path $\Gamma $ yields only one graph $G$ but there are
possibly many $\ n$--step paths covering each edge $\left(
b_{i-1}b_{i}\right) $ of $G$ exactly once which starts at $b_{1}$ and ends
at $b_{n}$. If we denote by $[\Gamma ]_{G}$ the set of all paths $\Gamma $
satisfying these conditions for a given admissible graph $G$, we have

\begin{proposition}
\label{TAG} Equation (\ref{Theta}) can be written as 
\begin{equation}
\Theta _{L}=\sum_{n\geq 1}\sum_{{{G:|E|=n-1}}\atop {{\rm admissible}}
}A_{G}\,W_{G}  \, , \label{Theta1}
\end{equation}
with
\begin{equation}
A_{G}:=\sum_{\Gamma \in \lbrack \Gamma \rbrack _{G}}\sum_{\Gamma =(\Gamma
_{1},\dots ,\Gamma _{s})}(-1)^{s+1}\,\bar{\alpha}_{\Gamma _{1}}\,\cdots \,
\bar{\alpha}_{\Gamma _{s}}  \, , \label{AG}
\end{equation}
where we sum over all sizes $n\in {\Bbb N},n\geq 1$, all 
admissible graphs $G$ of size ${|E|=n}$, over all paths $\Gamma $ in $
[\Gamma ]_{G}$ and over all decompositions of $\Gamma $ into $s,s\geq
1$, successive paths $(\Gamma _{1},\dots ,\Gamma _{s})$, each of which
capable of generating admissible graphs $G_{i}$. Here, with the
notation of (\ref{WG}) and footnote in Remark \ref{log}, 
\begin{equation}
(\!(v,W_{G}\,u)\!)=\vartheta _{b_{1}}\left( \prod_{\langle bb^{\prime
}\rangle \in E(G)}\Phi _{b,b^{\prime }}\right) \,\nu _{b_{n}} \, , \label{W-G}
\end{equation}
for $n>1$ with  $\Phi _{b_{1},b_{1}} =1$ for $n=1$ (the case that $G$ is
the trivial graph $(\{b_{1} \}, \emptyset)$).
\end{proposition}

We shall in the sequel state two lemmas and prove Theorem
\ref{difusion-k} under an extra assumption.

\begin{lemma}
\label{AG=0}If $G$ is an admissible graph with at least one cut--set contained
one edge (i.e. $G$ falls into two components by cutting a single
edge), then $A_{G}=0$. 
\end{lemma}

\begin{lemma}
\label{s-WG} There exist a constant $C_{[G]}<\infty $, depending on the
equivalence class $[G]$ of isomorphic graphs $G$, such that 
\begin{equation}
\sum_{{{G^{\prime }:G^{\prime }\sim G}}\atop {b_{1},b_{n}\,{\rm fixed}}
}\left| \prod_{\langle bb^{\prime }\rangle \in E(G^{\prime })}\Phi
_{b,b^{\prime }}\right| \leq \frac{C_{[G]}}{\left[ 1+{\rm dist}(b_{1},b_{n})
\right] ^{2d}}  \label{Prod}
\end{equation}
holds uniformly in $L$ for all admissible graph $G$ with $|E|=n$ 
and cut--sets with no less than two elements.
\end{lemma}

\begin{remark}
\label{GGG} The proof of Lemmas \ref{AG=0} and \ref{s-WG} are essentially
given in \cite{AKS} (see Assertions I and II of Section 4). Note that our
estimate (\ref{Prod}) have not included the logarithmic corrections
which appears in that reference. To get rid of these one has to control the
loop subgraphs of $G$ carefully as it is done in the ref. \cite{PPNM}. The
uniform upper bound (\ref{Prod}) results from the hypothesis that $G$
remains connected by cutting one single edge. 
\end{remark}

Graphs with single edge cut--sets do not contribute to (\ref{Theta1}) due to
the following cancellation in Lemma \ref{AG=0}.

\noindent
{\bf Proof of Lemma \ref{AG=0}.}
Let $\left( b_{i}\,b_{i+1}\right) $ be the only edge of a cut--set and let $
\Gamma =(\Gamma _{1},\dots ,\Gamma _{s})$ be a decomposition of a path in
$G$. Either  both  $b_{i}$ and $b_{i+1}$ belongs to some $\Gamma _{j}$ or
they belong to two successive ones. We call the latter decomposition type $A$
and the former type $B$.  It turns out that there is an one--to--one
correspondence between type 
$A$ and type $B$ decompositions differing only by the splitting of $\Gamma
_{j}$ into two elements $\Gamma _{j}^{(1)}$ and $\Gamma _{j}^{(2)}$. Lemma
\ref{AG=0} follows from the fact that the contribution to (\ref{AG}) of
a pair of decompositions established by this correspondence have the same 
absolute value with opposite signals. Note $\bar{\alpha}_{\Gamma
_{j}}=\bar{\alpha}_{\Gamma _{j}^{(1)}}\bar{\alpha}_{\Gamma
_{j}^{(2)}}$ if the edge $\left( b_{i}\,b_{i+1}\right) $ bridges the
two paths $\Gamma _{j}^{(1)}$ and $\Gamma _{j}^{(2)}$.

\hfill $\Box $

In view of Proposition \ref{TAG} and Lemmas \ref{AG=0} and \ref{s-WG},
equation (\ref{Theta}) can be estimated as 
\begin{equation}
|(\!(v,\Theta _{L}\,u)\!)|\leq \sum_{b,b^{\prime }\in {\Bbb B}_{L}}\vartheta
_{b}\,\frac{K_{L}}{\left[ 1+{\rm dist}(b,b^{\prime })\right] ^{2d}}\,\nu
_{b^{\prime }}  \, , \label{ThTh}
\end{equation}
where
\begin{equation}
K_{L}=\dsum_{n\geq 1}\dsum_{{[G]:|E|=n-1}\atop {{\rm admissible}}}A_{\left[ G
\right] }\,C_{[G]}  \, , \label{KL}
\end{equation}
with the sum running over the equivalence classes $[G]$ of isomorphic
admissible graphs $G$ of size $|E|=n$ and $C_{[G]}$ as in Lemma
\ref{s-WG}. Note that $A_{G}= A_{[G]}$.

Now we show that, if one uses, as in refs. \cite{AKS} and \cite{PPNM}, the
upper bound 
\begin{equation}\label{H1}
C_{[G]} \leq C^{r}
\end{equation}
for some geometric constant $C<\infty $ where $r=|V|$ is the number of
vertices in $G$, the equation (\ref{KL}) cannot be bounded uniformly in $L$. 
Taking into account property (\ref{alpha}),   
\begin{equation}\label{alphagamma}
|\bar{\alpha }_{\Gamma _{1}} \, \cdots \, \bar{\alpha } _{\Gamma _{s}} |
\leq \delta ^{n} 
\end{equation}

holds uniformly in $\Gamma $ and equation (\ref{KL}) can be bounded by 
\begin{equation}  \label{KL1}
K_{L} \leq \dsum _{n\geq 1} 
\left({{2n -1 }\atop {n }} \right) \, \delta ^{n} \sum _{r=1}^{{\rm min} \,
(n, |\Lambda |)} \Pi (n, r) \, C^{r} \,.
\end{equation}
Here, we have identified each path $\Gamma =\{b_{1},\dots ,b_{n}\} $ in a given
graph $G= (V,E)$ of size $|E|=n-1$ with a partition $P = (P _{1}, \dots ,P _{r})$
of $\{1, 2, \dots ,n  \}$ into $r=|V|$ pairwise disjoint subsets. This
association is one--to--one since $\Gamma $ is an ordered set of elements. 
Note that each component $P_{j}$ corresponds to one vertex $b_{i_{1}}
=b_{i_{2}}= \cdots = b_{i_{p}} $ of $G$ visited $p=|P_{j}|$ times by the path
$\Gamma $. One can thus replace the  
sum over all equivalent classes of graphs $[G]$ and over all paths $\Gamma $
in $[G]$ by the sum over all partitions $P$. The factor $\Pi (n, r)$ counts
the number of partitions of $\{1, 2, \dots , n\}$ into $r$ subsets. The bound
(\ref{KL1}) 
disregards the fact that $G$ is an admissible graph. Also, the consistency of
each decomposition $\Gamma = (\Gamma _{1}, \dots , \Gamma _{s})$ into paths
$\Gamma _{i}$'s which gives  rise to admissible graphs has been not considered.
The binomial factor $\left({{2n -1 }\atop {n }} \right)\leq 4^{n}$
counts the decomposition of $\Gamma $ with $n$ steps into any
number of paths with the number of steps $\leq n$ (the cardinality of the set
$\{1\leq i_{1} \leq i_{2}\leq \cdots  i_{n-1} \leq n-1\}$). In
addition, for the upper 
limit in the second sum we note that $r = |V|\leq (2 L-1)^{d}$ (the number of
vertices of $G$ cannot be larger than the number of sites in $\Lambda $).

Equation (\ref{KL1}) cannot be uniformly bounded since, from  the recursion
relation $\Pi (n, r) = \Pi (n-1, r-1) + r \, \Pi (n-1, r)$ (see \cite{W}), we
have
\[
\Pi (n, r) \geq r \Pi (n-1 , r) \geq r^{n-r} \Pi (r, r) = r^{n-r} \, ,
\]
which gives a factorial growth
\[
\sum _{r=1}^{{\rm min}  (n, |\Lambda |)} \Pi (n, r) \, C^{r} \geq 
\left\{
\begin{array}{lll}
(C\, n/2)^{n/2} & {\rm if } & n\leq |\Lambda |/2 \\
(C\, |\Lambda | /2)^{n - |\Lambda |/2} & {\rm if } & n > |\Lambda |/2
\, , 
\end{array} \right. 
\]
after replacing the sum by the term with $r={\rm min}  (n, |\Lambda |)/2$.

A sharper upper bound for (\ref{H1}) may be assumed if one think of  $C_{[G]}$
as being given by
\begin{equation}\label{C[G]}
C_{[G]}= \sup _{G' \sim G } \left| \prod_{\langle bb^{\prime }\rangle \in
E(G^{\prime })}\Phi _{b,b^{\prime }}\right| \left( 1+{\rm dist}(b_{1},b_{n})
\right) ^{2d} \, .
\end{equation}
As one varies the partition $P$ of $\{1, \dots , n \}$, the graph $G$, and the path
$\Gamma $ over it, varies accordingly and the decay of $\Phi _{b,b'}$ in this
formula can be useful. We propose that $C_{[G]}=C_{n,r}$ depends on the number
of vertices $r=|V|$ and edges $n=|E|$ as follows.

\begin{conjecture} \label{conj}
Let ${\widetilde{\Pi }} (n,r)= C_{n,r}\, \Pi (n,r)$. There exist a geometric
constant $C < \infty $ such that  
\begin{equation}\label{CPi}
{\widetilde{\Pi }} (n,r) \leq {\widetilde{\Pi }} (n-1,r-1) + C \, {\widetilde{\Pi
}} (n-1,r)  
\end{equation}
holds for $n,r \in \Bbb{N}$, $n\geq r$ with ${\widetilde{\Pi }} (r,r)=C^{r}$.
\end{conjecture}

Note that $\Pi (n,r)$ satisfies (\ref{CPi}) with $C$ replaced by $r$. Assuming
(\ref{CPi}) and using that ${\widetilde{\Pi }} (1,1)=C$ and ${\widetilde{\Pi
}} (k,l)=0$ if $k< l$, we have 
\begin{equation}\label{SPi}
\begin{array}{lll}
\dsum _{r=1}^{n} {\widetilde{\Pi }} (n,r) & \leq  & \dsum _{r=2}^{n}
{\widetilde{\Pi }} (n-1,r-1) + C \, \dsum _{r=1}^{n-1} {\widetilde{\Pi }}
(n-1,r) \\
 & = & (1+C) \dsum _{r=1}^{n-1} {\widetilde{\Pi }} (n-1,r) \\
& \leq & C(1+C)^{n} \, ,
\end{array}
\end{equation}
which leads (\ref{KL1}) to be bounded by
\begin{equation}\label{K_L}
K_{L} \leq \dsum _{n\geq 1} (4\delta )^{n} \dsum _{r=1}^{{\rm min}
(n,|\Lambda |)} {\widetilde{\Pi }} (n,r)  \leq  \dfrac {C \delta
'}{1-\delta '} \, ,
\end{equation}
where $\delta ' = 4 (1+C)\delta $.

This concludes the preliminaries and we are now ready to prove Theorem \ref
{difusion-k}. We observe at this point that no restrictions about the random
variables $\alpha _{b}$'s has been made beside (\ref{Ealpha}) and
(\ref{alphagamma}) with $\delta $ small enough. Has Conjecture
\ref{conj} been proved 
one could work along similar expansions to show that $\,(-\Delta _{L, \overline{
w}})^{-1/2}\left( -\Delta _{L, w}\right) \,(-\Delta _{L,
\overline{w}})^{-1/2}$ converges to $\left[ {\Bbb E}\,\left(
I-D_{L, w}\right) ^{-1}  \right] ^{-1}$ with probability $1$.

\medskip

\noindent {\it Proof of the upper bound of \ref{uplo}.} Let us recall some
facts about the matrix $-\Delta _{L, w}$. By equation (\ref{positive}) it is
a positive definite matrix and its square root is
well defined. We also have ${\Bbb E}\left( -\Delta _{L, w}\right)
=-\Delta _{L, \overline{w}}=-\overline{w}\,\Delta _{L}$ and, by Lemma
\ref{Green}, $ i(-\Delta _{L, w})^{-1}i^{\dagger }/L^{2}$ converges
with probability $1$ to $ (-\partial ^{2} (\kappa ))^{-1}$ exactly as
$i(-\nabla_{L} \cdot \kappa \nabla_{L})^{-1}i^{\dagger }/L^{2}$ does.

In view of this, we can apply Schwarz inequality to the following
identity\footnote{
In the following, for any two matrices $A$ and $B$, $A\leq B$ means $
(u,A\,u)\leq (u,B\,u)$ for all vectors $u$.}: 
\[
I=\left( {\Bbb E}\,(-\Delta _{L, w})^{1/2}\,(-\Delta _{L, w})^{-1/2}\right)
^{2}\leq {\Bbb E}\,(-\Delta _{L, w})\,{\Bbb E}\,(-\Delta _{L, w})^{-1}
\]
in order to get 
\begin{equation}
\left( {\Bbb E}\,(-\Delta _{L, w})^{-1}\right) ^{-1}\leq -\overline{w}\Delta
_{L}  \, ,\label{sch}
\end{equation}
which implies $\kappa \leq \overline{w}$ and concludes our assertion. \hfill 
$\Box $

\noindent {\it Proof of the lower bound of \ref{uplo}. }From equations (\ref
{1-D}), (\ref{D-D}) and (\ref{D1D}), we have 
\begin{eqnarray}
\left( {\Bbb E}\,(-\Delta _{L, w})^{-1}\right) ^{-1} &=&-\overline{w}\Delta
_{L}-\overline{w}(-\Delta _{L})^{1/2}\Theta _{L}(-\Delta _{L})^{1/2}
\label{sch1} \\
&=&-\overline{w}\nabla ^{\ast }R_{L}\nabla   \, , \nonumber
\end{eqnarray}
where $R_{L}:{\Bbb R}^{{\Bbb B}_{L}}\longrightarrow {\Bbb R}^{{\Bbb B}_{L}}$
is a matrix whose elements, in view of (\ref{ThTh}) and (\ref{K_L}), are
bounded by 
\begin{equation}
\left( R_{L}\right) _{b,b^{\prime }}\geq \delta _{b,b^{\prime }}-\frac{K_{L}
}{\left[ 1+{\rm dist}(b,b^{\prime })\right] ^{2d}}  \label{RL}
\end{equation}
with $K_{L}\leq \delta ^{\prime }C/\left( 1-\delta ^{\prime }\right) $. Note
from equations (\ref{sch}) and (\ref{sch1}) that $\Theta _{L}$ is a positive
matrix.

Using the isometry operators (\ref{iso}) and (\ref{iso2}) and the fact that $
i^{\dagger }i$ is the identity matrix in ${\Bbb R}^{|\Lambda |}$, we have $
L^{2}\,i\left( {\Bbb E}\,(-\Delta _{L, w})^{-1}\right) ^{-1}i^{\dagger }=
\overline{w}\left( L\,\,i\nabla ^{\ast }i^{\dagger }\,\right) \left(
iR_{L}i^{\dagger }\right) \left( L\,\,i\nabla i^{\dagger }\right) $ with $
L\,i\nabla ^{\ast }i^{\dagger }$ and $L\,i\nabla i^{\dagger }$ converging in 
$L_{0}^{2}\left( {\cal D}\right) $ to the operator $\partial =\left(
\partial /\partial \xi _{1},\ldots ,\partial /\partial \xi _{d}\right) $. In
addition, we claim that the kernel of $i\left( I-R_{L}\right) i^{\dagger }$
converges in distribution, as $L\rightarrow \infty $, to the delta function $
\delta (\eta ,\xi )$ times a $d\times d$ matrix $\varrho $, since the matrix
elements of $R_{L}$ decay faster than $1/{\rm dist}(b,b^{\prime })^{d}$ as $
{\rm dist}(b,b^{\prime })\rightarrow \infty $. For this, note that $i\left(
I-R_{L}\right) i^{\dagger }(\eta ,\xi )={\cal O}\left( L^{-d}\right) $ if $
\eta \neq \xi $ and $={\cal O}\left( L^{d}\right) $ if $\eta =\xi $. 

Whether $\varrho $ is a diagonal matrix cannot be decided by our estimates.
The results from this section leads to $\left( iR_{L}i^{\dagger }\right)
_{i,j}(\eta ,\xi )\longrightarrow \left( \delta _{i,j}+\varrho _{i,j}\right)
\delta (\eta ,\xi )$ and this implies  
\[
\kappa =\overline{w}\left( 1-\varrho \right) 
\]
where $1$ is the $d\times d$ identity and $\varrho $ is a positive matrix
satisfying $\varrho \leq \delta ^{\prime }C/\left( 1-\delta ^{\prime
}\right) $.

\hfill $\Box $

\medskip

\centerline{\bf Acknowledgment}
We wish to thank Luiz R. Fontes for helpful
discussions. R. da Silva was supported by CNPq under the PIBIC project
and D.H.U. Marchetti was partially supported by FAPESP and CNPq.


\end{document}